\let\latexaddtocontents\addtocontents
\documentclass[a4paper, twocolumn, 11pt, accepted=2024-04-30]{quantumarticle}
\let\addtocontents\latexaddtocontents

\pdfoutput=1
\usepackage[english]{babel}
\usepackage[utf8]{inputenc}
\usepackage[T1]{fontenc}

\usepackage[colorlinks=True, citecolor=blue, urlcolor=blue]{hyperref}
\usepackage{tikz}
\usetikzlibrary{quantikz}

\usepackage[backend=biber, citestyle=numeric-comp, sorting=none, doi=true, sortcites=true, maxnames=5]{biblatex}
\usepackage{amsmath,amssymb,amsfonts,physics,bbm}

\usepackage{graphicx}
\usepackage{bm}
\usepackage{xcolor}
\usepackage{bbold}
\usepackage{bbm}
\usepackage{float}
\usepackage{subfig}

\DeclareMathOperator*{\deph}{deph}
\DeclareMathOperator{\E}{\mathbb{E}}

\begin{document}
\title{Virtual mitigation of coherent non-adiabatic transitions by echo verification}
\author{Benjamin~F.~Schiffer}%
\email{benjamin.schiffer@mpq.mpg.de}
\affiliation{Max-Planck-Institut f\"ur Quantenoptik, Hans-Kopfermann-Str.~1, D-85748 Garching, Germany}%
\orcid{0000-0001-8951-2157}
\author{Dyon~van~Vreumingen}%
\email{d.vanvreumingen@uva.nl}
\affiliation{Institute of Physics, University of Amsterdam, Science Park 904, 1098 XH Amsterdam, The Netherlands}%
\affiliation{QuSoft, Centrum Wiskunde \& Informatica (CWI), Science Park 123, 1098 XG Amsterdam, The Netherlands}%
\orcid{0000-0003-2671-4434}
\author{Jordi~Tura}%
\email{tura@lorentz.leidenuniv.nl}
\affiliation{Instituut-Lorentz, Universiteit Leiden, P.O. Box 9506, 2300 RA Leiden, The Netherlands}%
\orcid{0000-0002-6123-1422}
\author{Stefano Polla}%
\email{polla@lorentz.leidenuniv.nl}
\affiliation{Instituut-Lorentz, Universiteit Leiden, P.O. Box 9506, 2300 RA Leiden, The Netherlands}%
\affiliation{Google Quantum AI, 80636 München, Germany}
\orcid{0000-0003-3909-0448}

\begin{abstract}
    Transitions out of the ground space limit the performance of quantum adiabatic algorithms, while hardware imperfections impose stringent limitations on the circuit depth. 
    We propose an adiabatic echo verification protocol which
    mitigates both coherent and incoherent errors, arising from non-adiabatic transitions and hardware noise, respectively. 
    Quasi-adiabatically evolving forward and backward allows for an echo-verified measurement of any observable. 
    In addition to mitigating hardware noise, our method uses positive-time dynamics only. 
    Crucially, the estimator bias of the observable is reduced when compared to standard adiabatic preparation, achieving up to a quadratic improvement.
\end{abstract}

\maketitle

\section{Introduction}
The study of quantum many-body systems requires the precise estimation of observables. Quantum state preparation is naturally a prerequisite to this end, which is the rationale behind quantum computers or quantum simulators.
The adiabatic algorithm has demonstrated large success in a variety of platforms~\cite{Barends2016Digitized, Ebadi2022Quantum}. 
Still, the performance of current devices is hindered by noise, which cannot be error corrected, yet. Therefore, error mitigation techniques have been explored both theoretically and experimentally and can significantly improve the estimation of observables~\cite{Cai2022Quantum, OBrien2022Purification, Kim2023Evidence}. 
Surprisingly, there have been few synergies jointly considering error mitigation for the adiabatic algorithm.

\

\begin{figure}[!t]
    \centering
    \includegraphics[width=1.0\linewidth]{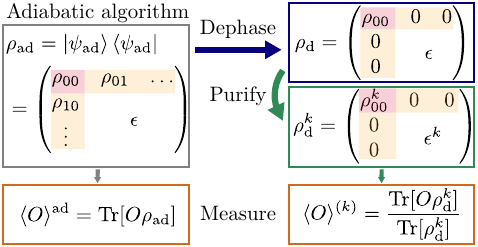}
    \caption{Schematic overview of the method. Density matrices are expressed in the energy eigenbasis of the target Hamiltonian. The pure state after the adiabatic evolution ($\rho_\text{ad}$) approximates the true ground state. Via a dephasing operation, the coherent error is promoted to an incoherent error in $\rho_\text{d}$ such that error mitigation techniques can be applied. This allows measuring the $k$\textsuperscript{th} degree purified observable $\langle O\rangle^{(k)}$ which yields a lower bias
    than evaluating the state directly after the adiabatic preparation $\left[\langle O\rangle^\text{ad}\right]$.}
    \label{fig:Fig1}
\end{figure} 

Any quantum circuit can be efficiently simulated by the adiabatic algorithm~\cite{Aharonov2007Adiabatic}. 
In adiabatic quantum computation, the system is initialized in the ground state of a trivial Hamiltonian and one seeks to prepare the ground state of the final Hamiltonian by slowly interpolating between the two. 
The success of the algorithm is determined by the speed of the adiabatic passage and spectral properties of the Hamiltonians~\cite{Messiah1962Quantum, Farhi2000Quantum}. 
More precisely, the total evolution time, or circuit depth, depends inverse polynomially on the minimum spectral gap between the ground state and the first excited state along the adiabatic path. 
These relations are quantified by the adiabatic theorem and versions thereof~\cite{Jansen2007Bounds, Amin2009Consistency, Wiebe2012Improved}. 
The adiabatic algorithm is especially suited for devices that implement dynamics natively without any Trotter overhead~\cite{Gross2017Quantum,Scholl2021Quantum, Bluvstein2022quantum, King2023Quantum}. 

To address the restrictions in current hardware, various error mitigation techniques have been explored in recent years to improve the usefulness of a noisy quantum computation~\cite{Cai2022Quantum}. 
These methods include zero-noise extrapolation, exploiting symmetry or purity constraints, and several other approaches. 
Here, we focus on purity methods, which aim to suppress stochastic errors by projecting the noisy state $\rho$ onto the closest pure state, given by the dominant eigenvector of $\rho$.

The purification can in general be achieved by collective measurements of several copies of $\rho$, known as virtual state distillation~\cite{Huggins_2021} or error suppression by derangement~\cite{Koczor2021Exponential}. 
Echo verification (EV) achieves this using two copies of $\rho$ multiplexed in time, rather than in space~\cite{Cai2021Resource,Huo2022Dual,OBrien2021Error}.
In EV, a desired state is prepared, an observable is measured controlled by an auxiliary qubit, and the state is then uncomputed. 
This allows to access expectation values of the so-called 2\textsuperscript{nd} degree purified state of $\rho$: $\expval{O}_\text{EV} = \Tr[O\rho^2]/\Tr[\rho^2]$. 
Recently, purification-based error mitigation has been tested experimentally in the context of the variational quantum eigensolver~\cite{OBrien2022Purification}.
Error mitigation methods tailored specifically to the adiabatic algorithm have been explored considerably less in the literature. 
Few exceptions consider error suppression and correction~\cite{Young2013Error} or symmetry-protection for Trotter dynamics~\cite{tran_faster_2021}.

\begin{figure}[t]
    \centering
       \includegraphics[width=\linewidth]{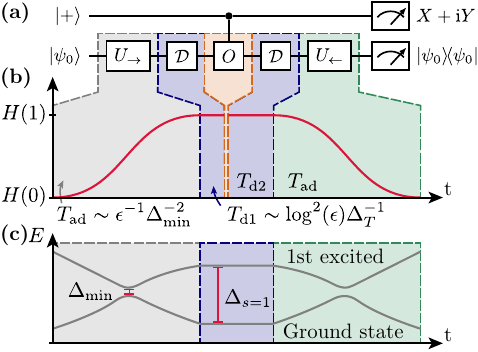}
    \caption{\textbf{(a)}~Quantum circuit for adiabatic echo verification to estimate an observable $\langle O\rangle$. A quasi-adiabatic sweep $U_\rightarrow$ is followed by an approximate ground state dephasing operation $\mathcal{D}$. After the controlled application of a unitary observable $O$ and dephasing again, the sweep is performed backward $U_\leftarrow (\neq U_\to^\dag)$.
    Postprocessing the measurement result, including the success information of the ground state projection, allows to extract an improved expectation value. \textbf{(b)}~Schematic of the Hamiltonian dynamics. 
    Approximate dephasing is implemented by evolving with the target Hamiltonian at $s=1$ for a random time. Typically, this time is much smaller than the time required for the adiabatic algorithm as depicted in \textbf{(c)}, where we sketch a corresponding low-energy spectrum.}
    \label{fig:Fig2}
\end{figure} 

In this work, we present a mitigation technique for estimating observables on quasi-adiabatically-prepared states, in the spirit of echo verification.
Along with stochastic device noise, our method seeks to suppress the coherent error due to non-adiabatic transitions.
Our method which we denote \emph{Adiabatic Echo Verification} (AEV) relies on dephasing operations to promote the coherent errors to random errors, which can then be mitigated. We illustrate the combination of dephasing and purification on a pure state in Fig.~\ref{fig:Fig1}. 
Similar to the original echo verification technique, the leading order error in the ground state expectation value of an observable is suppressed quadratically.
In particular, we consider an imperfect implementation of the dephasing operation using random-time evolution. 
Related random-time dynamics have been successfully used in the context of Zeno-type protocols~\cite{Boixo2009Eigenpath}.
The overhead from this dephasing operation is only poly-logarithmic in the accuracy of the dephasing operation for estimating observables of states within gapped phases. We discuss how the protocol compares favorably against doubling the total evolution time in the standard adiabatic algorithm. 
A key feature of our technique is that hardware noise is also mitigated naturally through the EV method.
Our protocol only requires implementing positive-time evolution and applying the operator of interest in a controlled way. Hence, the protocol is not only suitable for purely gate-based quantum devices but also for hybrid quantum simulators, e.g.~using neutral Rydberg atoms~\cite{Bluvstein2022quantum}.

\section{Background}
\subsection{The adiabatic algorithm}
In order to be able to measure observables on the ground state $\ket{E_0}$ of a target Hamiltonian $H_T$, a state approximating $\ket{E_0}$ with sufficient precision needs to be prepared.  
The quantum adiabatic algorithm (QAA) is a suitable algorithm for this task. 
At the heart of the QAA is the adiabatic theorem, which states that a system remains in an instantaneous eigenstate if the Hamiltonian is changed sufficiently slowly and the eigenstate is separated from other eigenstates by a minimum spectral gap $\Delta_\text{min}$ throughout the transition~\cite{Born1928Beweis}. 
Hence, the desired ground state $\ket{\psi_T}$ of a Hamiltonian of interest $H_T$ can be prepared by 
interpolating from a suitable Hamiltonian $H_0$ with a trivial ground state $\ket{\psi_0}$ as
\begin{align} \label{eq:adiabatic-hamiltonian}
    H(s)=(1-s) H_0 + s H_T.
\end{align}
where $s=t/T$ is the parametrized time. 
The folk version of the adiabatic theorem states that a total time $T = \mathcal{O}\left( \Delta_\text{min}^{-2}\epsilon^{-1/2} \right) $ suffices to prepare the ground state up to fidelity $1-\epsilon$. Rigorous versions of the adiabatic theorem give a bound $T = \mathcal{O}\left( \Delta^{-3}_\text{min}\epsilon^{-1/2}\right) $
if $H(s)$ is twice differentiable~\cite{Jansen2007Bounds, Amin2009Consistency}. 
Given a finite coherence time, the QAA prepares an approximation to the target state $\ket{\psi_\text{ad}} = \sqrt{1-\epsilon}\ket{E_0} + \sqrt{\epsilon} \ket{E_0^\perp}$ where $\langle E_0| E_0^\perp \rangle = 0$. Measuring an observable $O$, we obtain an approximation to the true value $\Tr\left[O\ketbra{\psi}_\text{ad}\right] = (1-\epsilon) \langle O \rangle_{\ket{E_0}} + \mathcal{O}(\sqrt{\epsilon})$.

\subsection{Purification-based error mitigation}

Purification methods such as echo verification (EV) or virtual state distillation improve the quality of an expectation value measurement on a noisy (incoherent) approximation $\rho$ of a pure state $\ketbra{\psi}$. 
This is achieved by effectively measuring the expectation value $\Tr[O\rho^k]$ of $O$ on the $k$-th power of the density matrix. 
Raising $\rho$ to the $k$-th power suppresses the eigenvectors with smaller eigenvalues, increasing the relative weight of the dominant eigenvector which, for small enough noise, should be $\ket\psi$.
As $\rho^k$ is non-normalized, purification methods prescribe to independently measure $\Tr[\rho^k]$ to calculate the desired estimator
\begin{equation} \label{eq:purification-estimator}
    \expval{O}^{(k)} \coloneqq \Tr[O\rho^k]/\Tr[\rho^k].
\end{equation}
If $\rho$ has an eigenstate $\ket{E_0}$ with large weight $c_0 = 1 - \epsilon$ (small positive $\epsilon$), we can write the density matrix as 
$ \rho = c_0 \ketbra{E_0} + \epsilon \rho_\perp $
with $\rho_\perp$ a density matrix orthogonal to $\ket{E_0}$ (i.e.,~$\rho_\perp\ket{E_0} = 0$).
The $k$\textsuperscript{th} degree purified estimator is then
\begin{align}
    \expval{O}^{(k)} &= \frac{c_0^k \bra{E_0}O\ket{E_0} + \epsilon^k \Tr[\rho_\perp^k O]}{c_0^k + \epsilon^k \Tr[\rho_\perp^k]} \\
    &= \bra{E_0}O\ket{E_0} + \mathcal{O}(\epsilon^k \Tr[\rho_\perp^k] \lVert O \rVert), \label{eq:EV}
\end{align}
where $\lVert \cdot \rVert$ is the operator norm.
Echo verification implements purification for $k=2$ using a single register by multiplexing two state-(un)preparation oracles in time. The method suppresses the error contributions to the observable estimator such that the leading order\footnote{
    In principle, the largest error contribution to $\Tr\left[O\ketbra{\psi}_\text{ad}\right]$ is of the order $\mathcal{O}(\sqrt{\epsilon})$ and is then mitigated to $\mathcal{O}(\epsilon^2)$. However, it would be misleading to interpret this as a general quartic suppression of the error. This is because the error contribution $\mathcal{O}(\sqrt{\epsilon})$ contains the term $\langle E_0| O | E_0^\perp\rangle$, which is often small independently of $\epsilon$. 
    This term can be rigorously sent to zero by applying a dephasing channel, which suppresses coherences in the Hamiltonian eigenbasis.
} 
becomes $\mathcal{O}(\epsilon^2)$.

\section{Mitigating coherent errors in adiabatic state preparation}
Our main contribution is to propose a method where the echo verification technique is applied to coherent errors. We focus on an application where the coherent error arises in the adiabatic algorithm due to finite algorithm runtimes. 
However, as the state prepared by a noiseless implementation of the adiabatic algorithm is pure, naive purification will not have any effect.

To recover the error mitigation power on $\rho_\text{ad} = \ketbra{\psi_\text{ad}}$, we introduce an ideal dephasing channel that turns coherent errors into incoherent noise, 
\begin{equation} \label{eq:ideal-dephasing}
    \deph_H[\rho] \coloneqq \sum_j \ketbra{E_j}{E_j} \rho \ketbra{E_j}{E_j} = \text{diag}[\rho],
\end{equation}
where we sum over an eigenbasis $\{\ket{E_j}\}_j$ of the target Hamiltonian $H_T$. Here, we assume a nondegenerate spectrum and give an extension for degenerate spectra in Appendix~\ref{app:degenerate}.
The dephasing channel projects a density matrix onto its diagonal in the energy eigenbasis, removing the off-diagonal coherences.
Applying the channel to the state prepared by the adiabatic algorithm yields
\begin{align} \label{eq:dephased-adiabatic-state}
    \rho_\text{d} \coloneqq \deph_H[\rho_\text{ad}]
    &= c_0 \ketbra{E_0} + \epsilon \rho_\perp \\
    &=
    \begin{pmatrix}
        c_0 & 0 & \ldots & 0 \\
        0 &  &  & \\
        \vdots & & \epsilon \rho_{\perp} & \\
        0 &  &  & \\
    \end{pmatrix}
\end{align}
with $\rho_\perp = \sum_{j\neq0} \rho_{jj}^{} \epsilon^{-1} \ketbra{E_j}$.
Then, using the echo verification technique on the dephased state, which is a mixed state, we obtain the following result for the observable $O$:
\begin{align}
   \frac{\Tr[O \rho_\text{d}^k]}{\Tr[\rho_\text{d}^k]} 
   = 
    (1 - \gamma)
    \bra{E_0} O \ket{E_0} 
    + 
    \gamma
    \frac{\Tr[O \rho_\perp^k]}{\Tr[\rho_\perp^k]},
\end{align}
with 
\begin{align}
\gamma &= \left[1 + c_0^k/\left(\epsilon^k \Tr[\rho_\perp^k]\right)\right]^{-1} \\
&\sim \mathcal{O}\left(\epsilon^k \Tr[\rho_\perp^k]\right).
\end{align}

To implement echo verification, typically, an inverse pair of unitaries $(U_\to, U_\to^\dag)$ would be required~\cite{OBrien2021Error}. 
The quantum circuit that implements adiabatic echo verification for an operator $O$ and a depasing operation is shown in Fig.~\ref{fig:Fig2}(a).
A priori, the unpreparation $U_\to^\dag$ would use negative-time dynamics, which is generally not available in analog simulators.
For our purposes, however, we can consider the two states $\rho_\text{ad} = U_\to \ketbra{\psi_0} U_\to^\dag$ and $\sigma_\text{ad} = U_\leftarrow^\dag \ketbra{\psi_0} U_\leftarrow$, where $U_\leftarrow$ is a positive-time adiabatic evolution with an inverted schedule from $s=1$ to $s=0$.
Both states have the same guaranteed fidelity with the target state $\ket{E_0}$ from the adiabatic theorem and ground state coherences are surpressed after the dephasing operation. This allows to use positive-time dynamics for the unpreparation step in AEV. We illustrate the inverted schedule in Fig.~\ref{fig:Fig2}(b), where a red line indicates the Hamiltonian in the dynamics.

Next, we consider the implementation of the dephasing channel. Importantly, we observe that a channel that dephases only the ground state would also be sufficient to achieve our goal, producing a state of the form Eq.~\ref{eq:dephased-adiabatic-state} with a more general, non-diagonal $\rho_\perp$, provided that $c_0$ still dominates. 
In the following part, we analyze such an approximate dephasing operation using positive-time dynamics.

\subsection{Implementation and cost of the dephasing}
\label{sec:dephasing-implementation}
We can implement an approximation of the dephasing channel (Eq.~\ref{eq:ideal-dephasing}) by a random-time evolution $\exp(-i H_T \tau)$, with $\tau$ sampled from a probability distribution $P(\tau)$, as follows:
We limit the support of $P$ to the interval $\tau \in [0, T_\text{d}]$. This ensures the dephasing can be realized naturally in quantum simulators and limits the time overhead of the dephasing operation to $2\,T_{\text{d}}$ for the AEV circuit.
We define the approximate dephasing channel
\begin{align} \label{eq:approximate-dephasing}
    \deph_{H, P}[\rho] 
    &\coloneqq 
    \int_0^{T_\text{d}} \dd P(\tau) e^{-i H_T \tau} \rho e^{i H_T \tau}
    \\
    &= \sum_{j,k} \mathcal{F}_{jk} \ketbra{E_j} \rho \ketbra{E_k},
\end{align}
where $\mathcal{F}_{jk} \coloneqq \mathcal{F}[P](E_j - E_k)$ is the Fourier transform of the random-time distribution at the transition energies. We will make use of the shorthand $\mathcal{D}[\rho] \coloneqq \deph_{H, P}[\rho]$.
As we only need to dephase the ground state, we require $\max_{j>0} |\mathcal{F}_{0j}| < \delta$.
Evaluating the adiabatic echo verification circuit with the \emph{approximate} dephasing channel $\mathcal{D}[\rho]$ yields an estimator with expectation
\begin{equation}
    \expval{O}^\text{AEV} = \frac{\Tr[O \tilde{\rho} \tilde{\sigma}]}{\Tr[\tilde{\rho} \tilde{\sigma}]}
\end{equation}
where $\tilde{\rho}_{jk} =  \mathcal{F}_{jk} [\rho_\text{ad}]_{jk}$ and $\tilde{\sigma}_{kl} = \mathcal{F}_{kl}^* [\sigma_\text{ad}]_{kl}$, expressed as matrix elements in the eigenbasis of the target Hamiltonian (cf.~Appendix~\ref{app:approximate}).
We can bound the deviation of the AEV estimator from the ground state expectation value as
\begin{equation}
    \left|\expval{O}^\text{AEV} - \bra{E_0} O \ket{E_0}\right| 
    \lesssim 
    \| O \| (\epsilon^{1/2} \delta +  \epsilon^{2})
\end{equation}
with a small prefactor. 
To ensure this error is bounded by $\mathcal{O}(\epsilon^2)$, it is then sufficient to take $\delta \sim \epsilon^{3/2}$.

An upper bound on the $|\mathcal{F}_{0j}|$ can be obtained as a functional of the distribution $P(\tau)$. We can thus redefine
\begin{equation}
    \delta \coloneqq \max_{\Delta > \Delta_T} \big|\mathcal{F}[P](\Delta)\big|
\end{equation}
where $\Delta_T < E_1 - E_0$ is a lower bound on the target Hamiltonian ground state gap.
In principle, different distributions can be chosen.
We might, for example, simply choose a uniform distribution $P(\tau) = 1/T_\text{d}$ for $\tau \in [0, T_{\text{d}}]$. As its Fourier transform is the cardinal sine function $\sin(x)/x$, we obtain $\delta \sim (\Delta T_{\text{d}})^{-1}$.
However, discontinuities in $P$ or its derivatives limit the asymptotic decay of $\mathcal{F}[P]$ to a polynomial.
We can improve upon this without increasing the maximal evolution time by choosing a mollifier, i.e.~a smooth distribution supported on $[0, T_\text{d}]$. 
A suitable example for our purposes is the rescaled bump function
\begin{equation} \label{eq:rescaled-bump-function}
    P_{T_\text{d}}(\tau) 
    = 
    \begin{cases}
        \frac{2}{\mathcal{N} T_\text{d}}\exp\left[\frac{T_\text{d}^2}{4\tau(\tau - T_\text{d})}\right] \quad \text{if} \quad \tau \in [0, T_\text{d}], \\
        0 \quad \text{otherwise}, 
    \end{cases}
\end{equation}
where $\mathcal{N} \approx 2.25$ is a normalization factor.
The Fourier transform of this function decays super-polynomially. 
Adapting the results from Ref.~\cite{johnson2015}, we recover
\begin{equation}
    \delta <
    \sqrt{\frac{8 \pi}{\sqrt{e}}} \,
    {\left(T_\text{d} \Delta_T \right)^{-3/4}} \,
    \exp[-\sqrt{T_\text{d} \Delta_T / 2}],
\end{equation}
with the full derivation included in Appendix~\ref{app:dephase}. A dephasing time $T_{\text{d}} \sim \Delta_T^{-1} \log^2[\epsilon]$ is thus sufficient to achieve an overall error $\mathcal{O}(\epsilon^2)$. Often, one is interested in observables of states in gapped phases~\cite{Sachdev1999Quantum}, such that only the poly-logarithmic term contributes to a non-constant overhead. We include the dephasing time in the illustration in Fig.~\ref{fig:Fig2}(b). In Fig.~\ref{fig:Fig2}(c), we show a cartoon of a the lowest two eigenenergies of the instantaneous Hamiltonian $H(s)$, showing a small adiabatic gap $\Delta_\text{min}$ during the quasi-adiabatic sweep and a larger, possibly constant gap $\Delta_{s=1}$ of the final Hamiltonian $H(1)$.

\section{Comparison with standard adiabatic algorithm}
We seek to compare the method proposed here with the trivial alternative for improving the performance of the adiabatic algorithm, which simply consists of doubling the evolution time in the QAA.
In the standard adiabatic theorem, there is a polynomial relationship between the accuracy and the evolution time~\cite{Jansen2007Bounds}. 
In principle, the adiabatic theorem can be improved towards an exponential error dependence by assuming a sufficiently smooth schedule with vanishing derivatives at the beginning and end of the schedule~\cite{Wiebe2012Improved}. 
However, this is at the cost of passing the minimal spectral gap at a faster rate, which, in general, leads to more transitions. 

\begin{figure}[t]
    \centering
    \includegraphics[width=1.0\linewidth]{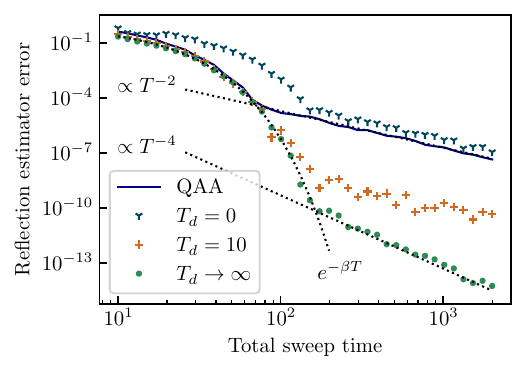}
    \caption{
        Comparing the QAA and AEV for different dephasing times $T_{\text{d}}$, as a function of the total sweep time $T$ (QAA: $T = T_\text{ad}$, AEV: $T = 2\,T_\text{ad}$). AEV improves over simply doubling the QAA sweep time in the regime of polynomial error dependence.
        The respective estimator bias from $\bra{E_0} O \ket{E_0}$ is shown, where $O = 1 - 2\ketbra{E_0}$ is a reflection on the target state.
        We perform density-matrix simulations; time-dependent evolution is implemented by Euler integration, ensuring a sufficiently small error when discretizing the sweep.
        Approximate dephasing is implemented with $P(\tau)$ as in Eq.~\ref{eq:rescaled-bump-function}.
    }
    \label{fig:Fig3}
\end{figure} 

Regarding our method, we therefore conclude that if the error dependence was indeed exponential, as in a Landau-Zener problem, the AEV would yield a performance comparable to the QAA with double the evolution time. Compared to the standard theorems with a polynomial dependence, our method improves up to quadratically. For the sake of concreteness, we include numerical benchmarks in Fig.~\ref{fig:Fig3} showing an advantage for preparing the ground state of a transverse field Ising model. 
We consider the reflection operator $O = 1 - 2\ketbra{E_0}$ on the ground state $\ket{E_0}$ and show the error from the expectation value $\bra{E_0} O \ket{E_0}$ as a function of the total sweep time. In Appendix~\ref{app:numerics}, we include additional numerics that consider the average magnetization of the state as another observable with practical relevance.
For the QAA, the sweep time is simply the time of one forward sweep. We compare this with the error we obtain using the AEV for three different scenarios: no dephasing ($T_d=0$), approximate dephasing ($T_d=10$) and perfect dephasing $T_d\rightarrow\infty$. Here, the total sweep time is the sum of the forward and the backward sweeps. The Hamiltonian chosen for the benchmark is an Ising model with a transverse and a longitudinal field, which is a non-integrable model. The quasi-adiabatic sweeps linearly interpolate from $H_0 = \sum_{j=1}^{5} \sigma^x_j$ to $H_T = 0.2 \sum_{j=1}^{5} \sigma^z_j - \sum_{j=1}^{4} \sigma^z_{j} \sigma^z_{j+1}$.
We observe an exponential scaling, in the so-called Landau-Zener regime, transitioning into an inverse-quadratic scaling for longer times (cf.~Ref.~\cite{Rezakhani2010Accuracy}). For these longer times, the AEV improves up to quadratically (error $\propto T^{-4}$) over the QAA (error $\propto T^{-2}$) if the detuning time is sufficiently large. The source code for the numerical simulations is published on Zenodo~\cite{adiab-mitig-code}.

\section{Discussion and practical considerations}
In this paper, we introduced \emph{Adiabatic Echo Verification} (AEV), a scheme to mitigate the coherent errors that characterize adiabatic state preparation. Our method is tailored to current quantum devices, which lack the possibility to correct errors. 
AEV requires doubling the circuit time compared to standard adiabatic state preparation, but improves up to quadratically in the estimator bias.
The additional features of the protocol are the following. First, in order to implement the verification part of the circuit, the path of the quasi-adiabatic evolution is simply reversed. Moreover, we show how the dephasing operation can be approximately implemented with positive-time dynamics. Hence, only positive-time evolution is required in AEV.
This makes our method suitable for quantum computers that operate in a hybrid mode of digital gates and analog simulation. Rydberg atom arrays have recently demonstrated such capabilities~\cite{Bluvstein2022quantum, Evered2023High}.

Additionally, AEV naturally mitigates non-coherent hardware noise through echo verification.
While our paper focuses on the theoretical analysis of the suppression of algorithmic noise, the hardware-error mitigation power of echo verification has been previously demonstrated in literature.
Numerical studies suggest that echo verification can effectively mitigate the effect of any single error, obtaining a quadratic improvement on the error on an observable from $\mathcal{O}(p)$ to $\mathcal{O}(p^2)$ with $p$ the error probability~\cite{OBrien2021Error}.
The effectiveness of echo verification for hardware error mitigation has also been demonstrated for a 10-qubit chemical simulation problem on a superconducting gate-based device~\cite{OBrien2022Purification}.

We note that our method is compatible with arbitrary sweep profiles in the QAA. This is especially helpful as it is well known that slowing down the adiabatic sweep at the position of the minimum spectral gap mitigates transitions out of the ground state~\cite{Roland2002Quantum, Schiffer_2022}.
More generally, our technique can be applied to other coherent approximate state preparation approaches, such as variational quantum algorithms (VQAs)~\cite{Cerezo2021Variational}. 
This applies to VQAs that prepare a pure state heuristically by a parametrized operation $U(\theta)$ aiming at approximating the desired ground state. 
By dephasing the prepared state and using the echo verification technique, unpreparing the state with $U(\theta)^{\dag}$, we expect that the performance of VQAs can be improved.

We note that the control-free versions of echo verification~\cite{OBrien2021Error,OBrien2022Purification}, which employ a reference state instead of a control qubit, are not naively available for AEV.
This is due the dephasing channel annihilating coherences between the reference state and the state of interest. 
Recently, a method for rescaling survival probabilities was considered that has similarities with control-free echo verification~\cite{Yang2023Simulating}. 
While their unnormalized estimator $\Tr[\rho O \rho O^{\dag}]$ differs from the echo verification counterpart $\Tr[O \rho^2]$, it would still allow for mitigating errors for certain interesting observables such as out-of-time-order correlators (OTOCs)~\cite{Mi2021}. 
Not requiring the implementation of a controlled operation can significantly simplify experiments. This is why an extension of AEV without a control qubit is an interesting direction for future work; a related idea has been recently proposed~\cite{shingu_quantum_2024}.
Another promising research direction is the combination of AEV with other purification-based error mitigation methods such as virtual state distillation ~\cite{Koczor2021Exponential, Huggins_2021}. Using multiple copies of the quasi-adiabatically prepared state, further improvements for surpressing errors seem possible. \\

\section*{Acknowledgements}
The authors thank A.~Barthe, T.~E.~O'Brien, E.~Koridon, M.~Malone, D.~Wild, and Y.~Yang for insightful discussions. 
This work received funding from the European Union’s Horizon 2020 research and innovation program under Grant No.~899354 (FET Open SuperQuLAN) and through the ERC StG FINE-TEA-SQUAD (Grant No. 101040729).  This research is part of the Munich Quantum Valley, which is supported by the Bavarian state government with funds from the Hightech Agenda Bayern Plus. J.T.~and D.vV.~acknowledge support from the Dutch Ministry of Economic Affairs and Climate Policy (EZK), through the Quantum Delta NL programme. This publication is also part of the ‘‘Quantum Inspire – the Dutch Quantum Computer in the Cloud’’ project (with project number [NWA.1292.19.194]) of the NWA research program ‘‘Research on Routes by Consortia (ORC)’’, which is funded by the Netherlands Organization for Scientific Research (NWO). S.P.~acknowledges support from Shell Global Solutions BV. J.T.~thanks the Google Research Scholar Program for support. We thank the Centro de Ciencias de Benasque Pedro Pascual for their hospitality. The views and opinions expressed here are solely those of the authors and do not necessarily reflect those of the funding agencies. Neither of the funding agencies can be held responsible for them.

\onecolumn
\printbibliography

\appendix

\section{Dephasing operation on a degenerate spectrum}\label{app:degenerate}

In Eq.~\ref{eq:ideal-dephasing}, we define the perfect dephasing channel for an operator $H$ with a non-degenerate spectrum.
If the spectrum of $H$ contains degeneracies, the dephasing channel will project $\rho$ to a block-diagonal operator, where the blocks are defined by the (degenerate) eigenspaces of $H$:
\begin{equation}
    \label{eq:degenerate-dephasing}
    \deph_H[\rho] = \sum_j \Pi_j \rho \Pi_j
\end{equation}
where $\Pi_j = \delta(H - E_j)$ is the projector on the eigenspace of $H$ with eigenvalue $E_j$.

For AEV, we are only interested in dephasing the ground state with respect to the rest of the spectrum; we only require the ground state of $H$ to be non-degenerate for Eq.~\ref{eq:dephased-adiabatic-state} and the subsequent analysis to be valid.
This is anyway a typical requirement in adiabatic state preparation.

\section{Evaluation of AEV estimator with approximate dephasing}\label{app:approximate}

In this section, we evaluate the error on the Adiabatic Echo Verification (AEV) estimator with respect to the target value $\bra{E_0} O \ket{E_0}$. We bound it as a function of the adiabatic state (un)preparation error $\epsilon$ and the dephasing approximation error $\delta$.
The AEV circuits we consider only require positive-time evolution with respect to the adiabatic Hamiltonian Eq.~\ref{eq:adiabatic-hamiltonian}, and the ability to perform a controlled-$O$ operation (or a decomposition thereof) to implement the Hadamard test.

We recall the echo verification (EV) expectation value estimator \cite{Polla2022Optimizing, OBrien2022Purification} is defined as 
\begin{equation}
    \expval{O}^\text{EV} \coloneqq \frac{\E[\text{verified Hadamard test circuit (VHT)}]}{\E[\text{echo circuit (Echo)}]}.
\end{equation}
In our case, the verified Hadamard test circuit is
\begin{equation} \label{eq:vht-circuit}
    \text{VHT} 
    \coloneqq
    \begin{tikzpicture}[baseline={([yshift=-.5ex]current bounding box.center)}]
        \node[scale=1] {
            \begin{quantikz}[row sep=0.1cm, column sep=0.32cm]
                \lstick{$\ket{+}$}  & \qw & \qw & \ctrl{1} & \qw & \qw & \qw & \meter{}\rstick{$X + \mathrm{i}Y$} \\
                \lstick{$\ket{\psi_0}$}  &  \gate{U_\rightarrow} &  \gate{\mathcal{D}} & \gate{O} & \gate{\mathcal{D}} & \gate{U_\leftarrow} & \qw & \meter{} \rstick{$\ketbra{\psi_0}$} &
            \end{quantikz}
        };
    \end{tikzpicture},
\end{equation}
where $U_\rightarrow$ is the adiabatic state preparation, $U_\leftarrow$ is the state unpreparation, and $\mathcal{D}\coloneqq\deph_{H, P}$ is the approximate dephasing. 
At the end of the circuit we need to measure $\mathrm{X} \otimes \ketbra{\psi_0}$ and $Y \otimes \ketbra{\psi_0}$ to recover the real $\Re[\expval{O}^\text{EV}]$ and imaginary $\mathrm{i}\Im[\expval{O}^\text{EV}]$ parts of the expectation value, respectively.
The output of a single sample of the circuit will be the result of the Pauli $X$ ($\pm1$) or $Y$ ($\pm \mathrm{i}$) on the control qubit if the system register returns to the state $\ket{\psi_0}$, and $0$ otherwise.
Our notation supposes that $O$ is a unitary operator, and its application is controlled by the state of the control qubit.
If $O$ is not unitary, we can rewrite it as a decomposition $O = \sum_x a_x \Re[U_x] + b_x \Im[U_x]$ and measure the terms of the decomposition separately~\cite{Polla2022Optimizing}.
The echo circuit, which is used to compute the normalization of $\expval{O}^\text{EV}$, is given as 
\begin{equation} \label{eq:echo-circuit}
    \text{Echo} \coloneqq
    \begin{tikzpicture}[baseline={([yshift=-.5ex]current bounding box.center)}]
        \node[scale=1] {
            \begin{quantikz}[row sep=0.1cm, column sep=0.32cm]
                \lstick{$\ket{\psi_0}$}  &  \gate{U_\rightarrow} &  \gate{\mathcal{D}} & \gate{\mathcal{D}} & \gate{U_\leftarrow} & \qw & \meter{} \rstick{$\ketbra{\psi_0}$} &
            \end{quantikz}
        };
    \end{tikzpicture}
\end{equation}
and obtained by substituting the operator $O$ with the identity in the previous circuit.

The adiabatic preparation and the adiabatic unpreparation are defined as
\begin{equation}
    U_\to = \mathcal{T}\exp{-\mathrm{i} \int_{0}^{T_\text{ad}} H[s(t)]  \, \dd t}
    \,; \quad
    U_\leftarrow = \mathcal{T}\exp{-\mathrm{i} \int_{0}^{T_\text{ad}} H[s(T_\text{ad} - t)] \, \dd t},
\end{equation}
where $\mathcal{T}\exp$ notates the time-ordered exponential, $H(s)$ is the adiabatic Hamiltonian (Eq.~\ref{eq:adiabatic-hamiltonian}), $s(t)$ the adiabatic schedule with $s(0) = 0$ and $s(T_\text{ad}) = 1$, and $T_\text{ad}$ is the total evolution time of the adiabatic algorithm.
Note that $\dd t$ is always positive, thus negative-time evolution is not required to implement $U_\to$ and $U_\leftarrow$. Typically, in an EV circuit, if the preparation unitary is $U$, then the unpreparation is performed with its conjugate transpose $U^\dag$ such that $UU^\dag = \mathbb{1}$. Here, however, this is not the case for the two operations $U_\leftarrow$ and $U_\rightarrow$. We will show that the added dephasing indeed removes this requirement for our purposes. 

In our calculations, we only assume that $U_\to$ ($U_\leftarrow$) implement an approximate state (un)preparation of $\ket{E_0}$ with a fidelity of at least $1 - \epsilon$ with small $\epsilon > 0$. Concretely, we define
\begin{equation}
    | \bra{E_0} U_\to \ket{\psi_0} |^2 = 1 - \epsilon_\leftarrow
    \,, \quad
    |\bra{\psi_0} U_\leftarrow \ket{E_0}|^2 = 1 - \epsilon_\rightarrow,
\end{equation}
such that $\epsilon = \text{max}\{\epsilon_\leftarrow, \epsilon_\rightarrow\}$. It is reasonable to assume the two adiabatic processes will have a similar error, as any adiabatic theorem bounds both in the same way. 

The approximate dephasing channel, of the form Eq.~\ref{eq:approximate-dephasing}, is defined via a matrix of Fourier coefficients 
\begin{align}
    \mathcal{F}_{jk} \coloneqq \mathcal{F}[P](E_j - E_k) \in \mathbb{C}
    \, , \qquad
    \mathcal{D}[\rho]_{jk} = \mathcal{F}_{jk} \rho_{jk}
\end{align}
We denote $A_{jk} = \bra{E_j} A \ket{E_j}$ the matrix elements of an operator in the eigenbasis of $H_T$.

The only requirement on the dephasing channel is that the Fourier coefficients are bounded $\max_{j>0} |\mathcal{F}_{0j}| < \delta$, which imposes that the coherences between the ground state and any other eigenstate are suppressed by a factor smaller than $\delta$.
In the main text, we relate this factor to the dephasing time and to ground state gap of the target Hamiltonian.

The expectation value of circuit Eq.~\ref{eq:vht-circuit} is
\begin{align}
    \E[\text{VHT}] 
    &=
     \Tr \Big\{
        U_\leftarrow ~\mathcal{D} \! \Big[
            \mathrm{Ctrl}O  ~\mathcal{D} \! \big[
                U_\to (\ketbra{+} \otimes \ketbra{\psi_0}) U_\to^\dag
            \big] \mathrm{Ctrl}O^\dag
        \Big] U_\leftarrow^\dag 
        \, (\ketbra{\psi_0} \otimes 2\ketbra{0}{1})
    \Big\} 
    \nonumber\\[1.5mm] &=
    \Tr \Big\{
        \underbrace{U_\leftarrow^\dag \ketbra{\psi_0} U_\leftarrow}_{\sigma} \,
        ~\mathcal{D} \! \Big[
            O ~\mathcal{D} \! \big[
                \,\underbrace{U_\to \ketbra{\psi_0} U_\to^\dag}_{\rho}\,
            \big] 
        \Big]
    \Big\}
    \nonumber\\ &=
    \sum_{jk} \mathcal{F}_{jk}
    \Tr \Big\{
        U_\leftarrow^\dag \ketbra{\psi_0} U_\leftarrow \,
        ~\mathcal{D} \! \Big[
            O \ket{E_j}\!
                \,\underbrace{\bra{E_j} U_\to \ketbra{\psi_0} U_\to^\dag \ket{E_k}}_{\rho_{jk}}\,
            \! \bra{E_k}
        \Big]
    \Big\}
    \nonumber\\ &=
    \sum_{jkl} \mathcal{F}_{jk} \mathcal{F}_{lk}
        \sigma_{kl} \,
        O_{lj} \,
        \rho_{jk}
\end{align}
where we expand the dephasing channels, and we define the density matrices $\rho$ and $\sigma$, corresponding respectively to the pure states
 \begin{align}
    U_\to\ket{\psi_0} &= 
    \sqrt{1 - \epsilon} \ket{E_0} + \sqrt\epsilon \sum_{j>0} \alpha_j \ket{E_j}
    \,, \quad
    \sum_{j>0} | \alpha_j |^2 = 1;
    \\
    \bra{\psi_0} U_\leftarrow &= 
    \sqrt{1 - \epsilon} \bra{E_0} +  \sqrt\epsilon \sum_{j>0} \beta_j^* \bra{E_j}
    \,, \quad
    \sum_{j>0} | \beta_j |^2 = 1.
\end{align}
We can then absorb the dephasing coefficients into $\tilde{\rho}_{jk} =  \mathcal{F}_{jk} \rho_{jk}$ and $\tilde{\sigma}_{kl} = \mathcal{F}_{lk} \sigma_{kl} =  \mathcal{F}_{lk}^* \sigma_{kl}$, simplifying
\begin{align}
    \E[\text{VHT}] = \Tr[\tilde{\rho}\tilde{\sigma}O]
    \, , \quad
    \E[\text{Echo}] = \Tr[\tilde{\rho}\tilde{\sigma}]
    \, , \quad
    \expval{O}^\text{EV} = \frac{\Tr[\tilde{\rho}\tilde{\sigma}O]}{\Tr[\tilde{\rho}\tilde{\sigma}]}.
\end{align}
Comparing this result to the standard purification estimator Eq.~\ref{eq:purification-estimator}, we see that the $\rho^2$ is substituted by $\tilde{\rho}\tilde{\sigma}$. The explicit expression for this operator in the $H_{\rm T}$ eigenbasis is
\begin{align}
    \tilde\rho\tilde\sigma = & \Big[ (1 - \epsilon)^2 + \epsilon(1 - \epsilon) \sum_{j>0} \alpha_j^* \beta_j \mathcal F_{j0}^* \mathcal F_{0j} \Big] \ket{E_0}\bra{E_0} \nonumber\\
    & {} + \sqrt\epsilon\sqrt{1 - \epsilon} \sum_{j>0} \Big[ (1 - \epsilon) \alpha_j \mathcal F_{j0} + \epsilon \sum_{l>0} \alpha_j \alpha_l^* \beta_l \mathcal F_{jl}\mathcal F_{l0}^* \Big] \ket{E_j}\bra{E_0} \nonumber\\
    & {} + \sqrt\epsilon\sqrt{1 - \epsilon} \sum_{j>0} \Big[ (1 - \epsilon) \beta_j^* \mathcal F_{j0} + \epsilon \sum_{l>0} \beta_j^* \alpha_l^* \beta_l \mathcal F_{jl}\mathcal F_{l0}^* \Big] \ket{E_0}\bra{E_j} \nonumber\\
    & {} + \epsilon \sum_{j,k>0} \Big[ (1 - \epsilon) \alpha_j \beta_k^* \mathcal F_{j0}\mathcal F_{0k}^* + \epsilon \sum_{l>0} \alpha_j \alpha_l^* \beta_l \beta_k^* \mathcal F_{jl} \mathcal F_{lk}^* \Big] \ket{E_j}\bra{E_k} \\
    = & [\tilde{\rho}\tilde{\sigma}]_{00} \ket{E_0}\bra{E_0} + \sum_{j>0} [\tilde{\rho}\tilde{\sigma}]_{j0} \ket{E_j}\bra{E_0} + \sum_{j>0} [\tilde{\rho}\tilde{\sigma}]_{0j} \ket{E_0}\bra{E_j} + \sum_{j,k>0} [\tilde{\rho}\tilde{\sigma}]_{jk} \ket{E_j}\bra{E_k}.
\end{align}

We proceed by bounding the error of $\expval{O}^\text{EV}$ with respect to the target $\bra{E_0} O \ket{E_0}$,
\begin{equation} \label{eq:error-def}
    \text{error}
    \coloneqq
    \left| 
    \frac{\Tr[O \tilde{\rho}\tilde{\sigma}]}{\Tr[\tilde{\rho}\tilde{\sigma}]}
    - \bra{E_0} O \ket{E_0}
    \right|
    = 
    \big| \!
    \Tr[O \tilde{\rho}\tilde{\sigma}]
    - O_{00}\Tr[\tilde{\rho}\tilde{\sigma}]
    \big| 
    \cdot
    \big| \! 
    \Tr[\tilde{\rho}\tilde{\sigma}]
    \big|^{-1}. 
\end{equation}

In the $H_T$ eigenbasis, the relevant terms read
\begin{align}
    \tr[O \tilde{\rho}\tilde{\sigma}]
    & =
    [\tilde{\rho}\tilde{\sigma}]_{00} O_{00} + 
    \sum_{j>0} [\tilde{\rho}\tilde{\sigma}]_{j0} O_{0j} +
    \sum_{k>0} [\tilde{\rho}\tilde{\sigma}]_{0k} O_{k0} +
    \sum_{j,k>0} [\tilde{\rho}\tilde{\sigma}]_{jk} O_{kj},
    \\
    \tr[\tilde{\rho}\tilde{\sigma}]
    & =
    [\tilde{\rho}\tilde{\sigma}]_{00}
    + \sum_{j>0} 
    [\tilde{\rho}\tilde{\sigma}]_{jj}.
\end{align}
We focus first on bounding the first factor on the right-hand side of Eq.~\ref{eq:error-def},
\begin{equation}
    \big| \!
        \tr[O \tilde{\rho}\tilde{\sigma}]
        -  O_{00} \tr[\tilde{\rho}\tilde{\sigma}]
    \big| 
    =
    \big| \!
        \tr[(O - O_{00}) \tilde{\rho}\tilde{\sigma}]
    \big|.
\end{equation}
We separate this expression through a triangle inequality,
\begin{align}
    \big| \!
        \tr[O \tilde{\rho}\tilde{\sigma}]
        -  O_{00} \tr[\tilde{\rho}\tilde{\sigma}]
    \big| 
    & \leq 
        \bigg| \sum_{j>0} [\tilde{\rho}\tilde{\sigma}]_{j0} O_{0j} \bigg| +
        \bigg| \kern-0.1em \sum_{k>0} [\tilde{\rho}\tilde{\sigma}]_{0k} O_{k0} \bigg| \nonumber \\
        & \quad + \bigg| \kern-0.1em \sum_{j,k>0} [\tilde{\rho}\tilde{\sigma}]_{jk} O_{kj} -
        \sum_{j>0} [\tilde{\rho}\tilde{\sigma}]_{jj} O_{00} \bigg|.
    \label{eq:error-bound-first-step}
\end{align}
To bound the first term, we first use that, by the definition of the spectral norm,
\begin{align}
    \bigg| \kern-0.1em \sum_{j>0} [\tilde{\rho}\tilde{\sigma}]_{0j} O_{j0} \bigg|
    \leq \|O\| \cdot \|[\tilde{\rho}\tilde{\sigma}]_{\cdot0}\|
\end{align}
where $[\tilde{\rho}\tilde{\sigma}]_{\cdot0}$ is the vector with entries $[\tilde\rho\tilde\sigma]_{j0}$ for $j > 0$ (and zero for $j = 0$). Let us explicitly write out an upper bound on $\|[\tilde{\rho}\tilde{\sigma}]_{\cdot0}\|^2$:
\begin{align} \label{eq:error_contrib_ground}
    \|[\tilde{\rho}\tilde{\sigma}]_{\cdot0}\|^2 &= \epsilon(1 - \epsilon) \sum_{j>0} \bigg| (1 - \epsilon) \alpha_j \mathcal{F}_{j0} + \epsilon \sum_{l>0} \alpha_j \alpha_l^* \beta_l \mathcal{F}_{jl} \mathcal{F}_{l0}^* \bigg|^2 \nonumber\\
    & \leq \epsilon(1 - \epsilon) \sum_{j > 0} |\alpha_j|^2 \bigg( (1 - \epsilon) |\mathcal{F}_{j0}| + \epsilon \bigg| \! \sum_{l>0} \alpha_l^* \beta_l \mathcal{F}_{jl} \mathcal{F}_{l0}^* \bigg| \bigg)^2 \nonumber \\
    & \leq \epsilon(1 - \epsilon) \sum_{j > 0} |\alpha_j|^2 \bigg( (1 - \epsilon) |\mathcal{F}_{j0}| + \epsilon \sum_{l>0} |\alpha_l^*| \, |\beta_l| \, |\mathcal{F}_{jl}| \, |\mathcal{F}_{l0}^*| \bigg)^2 \nonumber \\
    & \leq \epsilon(1 - \epsilon) \|\vec\alpha\|^2 \big( (1 - \epsilon)\delta + \epsilon \kern0.08em \delta \kern0.08em \|\vec\alpha\| \|\vec\beta\| \big)^2 \nonumber \\
    & = \epsilon(1 - \epsilon) \delta^2
\end{align}
where we note that $\vec\alpha$ and $\vec\beta$ are normalized by definition, $\mathcal{F}_{jk} \leq 1$ and  $\max_{k>0}|\mathcal{F}_{0k}| = \delta$; Cauchy-Schwarz was used in the fourth line. Hence, the first term in Eq. \ref{eq:error-bound-first-step} is upper bounded by $\|O\| \sqrt\epsilon \sqrt{1 - \epsilon} \kern0.08em \delta$. The same bound applies to the second term $\big| \! \sum_{k>0} [\tilde{\rho}\tilde{\sigma}]_{k0} O_{0k} \big|$. \par
The last term of Eq.~\ref{eq:error-bound-first-step} can be rewritten as
\begin{equation}
    \sum_{j,k>0} [\tilde{\rho}\tilde{\sigma}]_{jk} O_{kj} -
    \sum_{j>0} [\tilde{\rho}\tilde{\sigma}]_{jj} O_{00}
    = 
    \tr[
        \Pi_> \, \tilde{\rho}\tilde{\sigma} \, \Pi_> (O - O_{00} \mathbb{1})
    ]
\end{equation}
where $\Pi_{>} = \mathbb{1} - \ketbra{E_0}$ is the projector on the subspace orthogonal to $\ket{E_0}$.
We can then use the Von Neumann inequality to bound
\begin{equation}
    \big| \! \tr[
        \Pi_> \, \tilde{\rho}\tilde{\sigma} \, \Pi_> (O - O_{00} \mathbb{1})
    ] \big|
    \leq
    \| O - O_{00} \mathbb{1} \| \cdot
    \| \Pi_> \, \tilde{\rho}\tilde{\sigma} \, \Pi_> \|_1.
\end{equation}

Now, by virtue of the triangle inequality, and the fact that $\|\ket u\bra v\|_1 = \|u\| \|v\|$, for any vectors $\ket u, \ket v$, we have 
\begin{align}
    \|\Pi_>\tilde\rho\tilde\sigma\Pi_>\|_1 &\leq \epsilon(1 - \epsilon) \, \bigg\| \sum_{j>0} \alpha_j \mathcal{F}_{j0} \ket{E_j} \bigg\| \, \bigg\| \sum_{k>0} \beta_k \mathcal{F}_{0k} \ket{E_k} \bigg\| \nonumber \\
    & \quad + \epsilon^2 \sum_{l>0} |\alpha_l| |\beta_l| \, \bigg\| \sum_{j>0} \alpha_j \mathcal{F}_{jl} \ket{E_j} \bigg\| \, \bigg\| \sum_{k>0} \beta_k \mathcal{F}_{lk} \ket{E_k} \bigg\| \nonumber\\
    &\leq \epsilon(1 - \epsilon)\delta^2 + \epsilon^2 \, (\max_{j,k>0}|\mathcal{F}_{jk}|)^2 \, \sum_{l>0} |\alpha_l| |\beta_l| \nonumber\\
    &\leq \epsilon(1 - \epsilon)\delta^2 + \epsilon^2 \label{eq:error_contrib_excited_bulk}
\end{align}
where we used Cauchy-Schwarz in the last line together with the normalisation of $\vec\alpha$ and $\vec\beta$.

Combining the bounds from Eqs. \eqref{eq:error_contrib_ground} and \eqref{eq:error_contrib_excited_bulk}, and using that $\| O - O_{00} \mathbb{1} \| \leq 2 \| O \|$, we get
\begin{equation}
    \big| \!
        \tr[O \tilde{\rho}\tilde{\sigma}]
        -  O_{00} \tr[\tilde{\rho}\tilde{\sigma}]
    \big| 
    \leq
    2 \| O \| \left[
        (1 - \epsilon)^{1/2}\, \epsilon^{1/2} \delta +
        \epsilon (1-\epsilon) \delta^2 + \epsilon^2 
    \right].
\end{equation}
Next, to bound the factor $| \! \tr[\tilde{\rho}\tilde{\sigma}] |^{-1}$ in Eq.~\ref{eq:error-def}, we apply the reverse triangle inequality to $| \! \tr[\tilde{\rho}\tilde{\sigma}] |$:
\begin{align}
    \big| \! \tr[\tilde\rho\tilde\sigma] \big| &= \bigg| (1 - \epsilon)^2 + 2\epsilon(1 - \epsilon) \Re \Big( \sum_{j>0} \alpha_j \beta_j^* \mathcal{F}_{j0}^2 \Big) + \epsilon^2 \sum_{j,l>0} \alpha_j \alpha_l^* \beta_l \beta_j^* \mathcal{F}_{jl}^2 \bigg| \nonumber\\
    &\geq \big| (1 - \epsilon)^2 - 2\epsilon(1 - \epsilon) - \epsilon^2 \big| \label{eq:tr_trho_tsigma_ineq}
\end{align}
where we used that $\delta \leq 1.$ For $\epsilon < \sqrt{3/2} - 1$, the argument in Eq.~\ref{eq:tr_trho_tsigma_ineq} is strictly positive, so we can remove the absolute value symbol. In this case, we have $\big| \! \tr[\tilde\rho\tilde\sigma] \big|^{-1} \in \mathcal O(1)$. The dominant terms in the error Eq.~\ref{eq:error-def} are then
\begin{equation}
    \text{error} \sim \| O \| (\epsilon^{1/2} \delta +  \epsilon^{2}).
\end{equation}
We can verify that for $\delta \to 0$ we recover the error scaling with $\epsilon^2$, as expected from perfect dephasing.
To achieve the same scaling, it is in fact sufficient to choose $\delta = \epsilon^{3/2}$.

\section{Dephasing time for a smooth probability distribution}\label{app:dephase}

In this section we motivate the choice of the rescaled bump function in Eq.~\ref{eq:rescaled-bump-function} for the distribution $P(\tau)$ used to implement dephasing by random-time evolution (Eq.~\ref{eq:approximate-dephasing}).
We recall that we require $P : [0, T_\text{d}] \to \mathbb{R}_+$ to have support on $[0, T_\text{d}]$.
This ensures we only need to evolve for positive times and the maximal dephasing time is $T_\text{d}$.
The performance for the dephasing operation on the ground state is measured by $\delta = \max_{\Delta > \Delta_T} | \mathcal{F}[P](\Delta) |$, which is essentially a bound on the decay of the Fourier transformation of $P$.
As $\tau$ and $\Delta$ are conjugate dimensionful variables, we can equivalently study
\begin{equation}
    \delta = \max_{\omega > \Delta_T T_{\text{d}}} | \mathcal{F}[\tilde{P}](\omega) |
    \quad \text{for} \quad
    \tilde{P} : [0, 1] \to \mathbb{R}_+,
\end{equation}
where $\tilde{P}(\tau\Delta_T) = P(\tau)$.

To obtain the best possible asymptotic decay of the Fourier transform of a function $\mathcal{F}[f]$, we should choose $f$ to be smooth.
In fact, requiring the Fourier transform of $f$ to decay as $\mathcal{F}[f](k) \lesssim |k|^{-(r+1+\epsilon)}$ (for any choice of $\epsilon>0$)
implies that
\begin{equation}
    \exists L_r>0: \forall x 
    \quad 
    \left\lvert \frac{\dd^r f(x)}{\dd x^r} \right\rvert
    = 
    \bigg\lvert \int \dd k \, e^{i k t} 
    \underbrace{k^r \mathcal{F}[f](k)}_{\lesssim |k|^{-(1+\epsilon)}}
    \bigg\rvert
    < L_r,
\end{equation}
because $|k|^{-(1+\epsilon)}$ is absolutely integrable away from 0 and $\mathcal{F}[f](k)$ is bounded.
This implies that $f$ and all its derivatives up to order $r-1$ are Lipschitz continuous.
Thus, to achieve a Fourier transform decaying faster than any polynomial $\mathcal{F}[f](k) = o(1/\text{poly}(k))$, we have to choose $f(x) \in C_\infty$ a smooth function.

One smooth function with compact support is the bump function
\begin{equation}
    f(x) = \begin{cases}
        e^{-(1 - x^2)^{-1}} \quad \text{if} \quad -1 < x < 1, \\
        0 \quad \text{otherwise}, 
    \end{cases}
\end{equation}
we define its norm $\mathcal{N} \coloneqq \int_{-1}^1 \,f(x) \,\dd x \approx 2.25$.
Based on this function, we define the probability distribution
\begin{equation}
    P_{T_\text{d}}(\tau) 
    =  \frac{2}{T_\text{d}^{}\mathcal{N}} f\left(2 \frac{\tau}{T_\text{d}} - 1\right)
    =
    \begin{cases}
        \frac{2}{T_\text{d}^{}\mathcal{N}} \exp\left(\left[4(\frac{\tau}{T_\text{d}}-1)\frac{\tau}{T_\text{d}}\right]^{-1}\right) \quad \text{if} \quad 0 < \tau < T_\text{d}, \\
        0 \quad \text{otherwise}, 
    \end{cases}
\end{equation}
which is normalized, smooth, and has support on $[0, T_\text{d}]$.
The Fourier transform of this function can be estimated through the saddle point approximation; we build on the results of Ref.~\cite{johnson2015} which provide a bound the Fourier transform $\mathcal{F}[\mathcal{N} f]$ of the normalized $f(x)$:
\begin{equation}
    \mathcal{F}[\mathcal{N} f](k) \approx 
    2 \Re\left[
        \sqrt{\frac{-i \pi}{\sqrt{2 i}}} e^{ik - \frac{1}{4} - i \sqrt{k}}
    \right] k^{-\frac{3}{4}} e^{-\sqrt{k}}.
\end{equation}
We construct a monotonic envelope for this oscillating function by substituting the real part for an absolute value, and we perform a change of variables obtaining the bound
\begin{equation}
    \delta = 
    \max_{\Delta > \Delta_T}|\mathcal{F}[P_{T_{\text{d}}}](\Delta)|
    <
    \sqrt{\frac{8 \pi}{\sqrt{e}}} \,
    {\left(T_\text{d} \Delta_T \right)^{-3/4}} \,
    e^{-\sqrt{T_\text{d} \Delta_T / 2}}.
\end{equation}
The validity of this bound is also verified numerically.
This translates to a statement on the dephasing time $T_{\text{d}}$ required to achieve a target dephasing performance $\delta_\text{target}$ for a given gap $\Delta_T$ between the ground state and the first excited state of $H_T$.
Note that the inverse is defined in terms of the principal branch $W_0$ of the Lambert $W$ function. We obtain
\begin{equation}
    T_\text{d} =
    \frac{9}{2 \, \Delta_T} W_0\left[
        \frac{2 \, \sqrt{2} \, \pi^{1/3}}{3 \, e^{1/6} \, \delta_\text{target}^{2/3}}
    \right]
    \lesssim 
    O\left(\log^2(\delta_\text{target}^{-1}) \Delta_T^{-1} \right).
\end{equation}
Thus, $T_\text{d}$ grows linearly with $\Delta_T$ and poly-logarithmically with $\delta_\text{target}^{-1}$.

\section{Numerical simulations for magnetization in an Ising model}\label{app:numerics}
To complement the numerics in the main text, we include additional simulations where we consider the total magnetization 
\begin{align}
    M = \sum_i^n \sigma_i^z 
\end{align}
as the observable in the AEV protocol.
The considered model is the same Ising model as for Fig.~\ref{fig:Fig3}, where the reflection along the target ground state was probed instead of the magnetization.
We observe a strong similarity in the behavior of the error between both observables.

\begin{figure}[t]
    \centering
    \includegraphics[width=.48\linewidth]{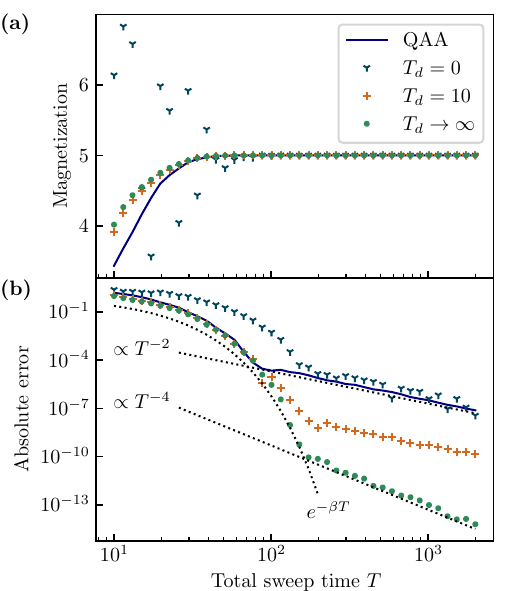}
    \caption{
        Numerical simulation of the AEV protocol on Ising chain of five spins; this is the same model as in Fig~\ref{fig:Fig3}, but here the measured observable is the total magnetization $M = \expval{\sum_{j=1}^5 \sigma_j^z}$.
        \textbf{(a)} Expectation values for the magnetization $M$, comparing the quantum adiabatic algorithm (QAA) with the AEV for different dephasing times. 
        While all scenarios converge asymptotically towards the true value for sufficiently long sweep times, they differ in their convergence behavior.
        \textbf{(b)} Absolute error of the different scenarios with the true value. The convergence of the magnetization behaves qualitatively as the reflection operation in Fig.~\ref{fig:Fig3}.
    }
    \label{fig:Fig4}
\end{figure}

Note that the error on the measured observable in AEV with approximate dephasing (see Sec.~\ref{sec:dephasing-implementation}) can be positive or negative, thus the results can overshoot variational bounds.
This is because the result of AEV is obtained as a fraction between expectation value of two circuits, either of which having independent variationally-bounded errors.
In practice, in Fig.~\ref{fig:Fig4} we observe a value of the magnetization overshooting the bound of $\expval{M} \leq 5$ only for $T_d = 0$. 
A modest amount of dephasing suppresses the error in the echo circuit, suppressing the oscillations in the sign of the error and recovering a variationally-bounded result.

\end{document}